\documentclass[12pt,a4paper]{article}
\begin{document}
\large

\newpage
\begin{center}
{\bf MASS STRUCTURE OF AXIAL-VECTOR TYPES
OF LEPTONS AND FIELDS}
\end{center}
\vspace{0.5cm}
\begin{center}
{\bf Rasulkhozha S. Sharafiddinov}
\end{center}
\vspace{0.5cm}
\begin{center}
{\bf Institute of Nuclear Physics, Uzbekistan Academy of Sciences,
Tashkent, 100214 Ulugbek, Uzbekistan}
\end{center}
\vspace{0.5cm}

A classification of currents with respect to C-operation admits the
existence of C-noninvariant types of Dirac fermions. Among them one
can meet an electroweakly charged C-antisymmetrical leptons, the mass
of which includes the electric and weak components responsible for the
existence of their anapole charge, charge radius and electric dipole moment.
Such connections can constitute the paraleptons of axial-vector currents, for
example, at the interactions with field of spinless nuclei of true neutrality.
We derive the united equations which relate the structural parts of mass to
anapole, charge radius and electric dipole of any truly neutral lepton in the
framework of flavour symmetry. Thereby, they establish the C-odd nature of
leptons and fields at the level of constancy law of the size implied from the
multiplication of a weak mass of C-antisymmetrical lepton by its electric mass.
Therefore, all leptons of C-antisymmetricality regardless of the difference in
masses of an axial-vector character, have the same anapole with his radius as
well as an equal electric dipole. Their analysis together with measured value
of an electric dipole moment of lepton gives the right to interpret not only
the existence of truly neutral types of leptons and fields but also the
availability of mass structure in them as the one of earlier laboratory facts.

\newpage
\begin{center}
{\bf 1. Introduction}
\end{center}

The question about the electric charge of C-noninvariant character is highly
important for elucidation of nature of mass of truly neutral fermions. This
charge regardless of whether it is hypercharge or anapole \cite{1}, may serve
as the source of an electric dipole \cite{2}.

Such a connection for the first time was established by the author \cite{3}
on the basis of flavour dynamics of elementary particles. However, at the
unification of the anapole and electric dipole there arises a range of
nontrivial problems connected with symmetry properties of these types of
currents. The point is that the anapole and electric dipole stating that the
same fermion may have simultaneously both T-symmetrical and T-antisymmetrical
nature are, respectively, the CP-even and CP-odd components of P-noninvariant
axial-vector $A_{l}$ current. From the point of view of any of particles, the
same lepton may not be simultaneously both a C-invariant and a C-noninvariant
fermion. A given circumstance becomes more interesting if we take into account
that to each type of charge corresponds a kind of the dipole \cite{4}. One
of such pairs expresses the CP-invariance as well as the C-noninvariance of
an electric dipole. This, however, can explain the absence of T-symmetry of
earlier T-invariant anapole even at the violation of CPT-parity of all
types of C-antisymmetrical $A_{l}$ currents \cite{5}.

It is not a wonder therefore that a classification of leptonic currents with
respect to C-operation admits the existence of C-noninvariant types of leptons
$(l^{A}=e^{A}, \mu^{A}, \tau^{A}, ...)$ possessing an individual flavours
\cite{5} which distinguish any of them from others by a true charge. It is
characterized by a quantum number conserved at all interactions between
C-antisymmetrical $(l^{A}=C\overline{l}^{A})$ fermion and field of emission
of an axial-vector nature analogously to the fact that lepton flavours or
full lepton number are conserved in the processes with well known C-invariant
$(l^{V}=-C\overline{l}^{V})$ leptons $(l^{V}=e^{V}, \mu^{V}, \tau^{V}, ...)$
of vector $V_{l}$ currents \cite{6,7}. Such a true flavour says in favor of
the unidenticality of truly neutral leptons and antileptons from the point
of view of CP-symmetricality $(l^{A}=-CP\overline{l}^{A})$ of these types
uf particles.

One can define the structure of their interaction with virtual axial-vector
photon \cite{9} in the limit of quantum electrodynamics by the
following manner:
\begin{equation}
j_{em}^{A_{l}}=\overline{u}(p',s')\gamma_{5}[\gamma_{\mu}
G_{1l^{A}}(q^{2})-i\sigma_{\mu\lambda}q_{\lambda}G_{2l^{A}}(q^{2})]u(p,s).
\label{1}
\end{equation}
Here $\sigma_{\mu\lambda}=[\gamma_{\mu},\gamma_{\lambda}]/2,$ $l=e,$ $\mu,$
$\tau,$ $p(s)$ and $p'(s')$ imply the four-momentum (helicities) of fermion
before and after the emission of an axial-vector character,
$q=p-p'$ is the momentum transfer.

The anapole and electric dipole form factors $G_{il^{A}}$ constitute,
respectively, the Dirac $(i=1)$ and Pauli $(i=2)$ components of leptonic
$A_{l}$ current \cite{3}. We present their in the form \cite{10}
\begin{equation}
G_{il^{A}}(q^{2})=g_{il^{A}}(0)+R_{il^{A}}(q^{2})+...
\label{2}
\end{equation}
where $g_{il^{A}}(0)$ define the static value of the anapole and electric
dipole moment of a particle. The finctions $R_{il^{A}}(q^{2})$ characterize
the lepton axial-vector radius dependence of currents.

The interaction between the anapole $r_{l^{A}}$ radius and field of emission
can be described by the form factor
$R_{1l^{A}}(q^{2})=(q^{2}/6)<r^{2}_{l^{A}}>.$

It is clear, however, that the same anapole may not be both a weak and
an electric charge of an axial-vector fermion. At the same time, the nature
itself has been created so that to each type of charge corresponds a kind
of inertial mass \cite{11}. Therefore, any lepton with an electroweak behavior
must have the weak \cite{12} as well as the electric mass. Unlike a C-invariant
particle, the mass of which is strictly a vector type, C-noninvariant leptons
have the mass of an axial-vector character \cite{3}.

Thus, it follows that all the axial-vector mass $m_{l^{A}}$ and charge
$e_{l^{A}}$ of an electroweakly charged C-antisymmetrical lepton coincide
with its electroweak $(EW)$ mass and charge
\begin{equation}
m_{l^{A}}=m_{l^{A}}^{EW}=m_{l^{A}}^{E}+m_{l^{A}}^{W},
\label{3}
\end{equation}
\begin{equation}
e_{l^{A}}=e_{l^{A}}^{EW}=e_{l^{A}}^{E}+e_{l^{A}}^{W}
\label{4}
\end{equation}
including the Coulomb $(E)$ and weak $(W)$ contributions.

They constitute an intralepton harmony of the four types of forces \cite{13}
which has the crucial value for steadiness of the anapole charge distribution
in lepton of true neutrality. In this appears a latent structure of truly
neutral lepton interaction with the Coulomb, weak and the electroweakly
interference fields of emission and of a fundamental role in its formation
of an axial-vector electroweak mass structural parts.

To show their features, we investigate here the nature of truly neutral
types of leptons and fields at the elastic scattering on a spinless nucleus
going as a result of Coulomb $(E)$ and weak $(W)$ masses and of the anapole
$g_{1l^{A}}(0)$ charge, charge $r_{l^{A}}$ radius and electric dipole
$g_{2l^{A}}(0)$ moment of longitudinal polarized fermions
of axial-vector weak neutral $A_{l}$ currents.

Insofar as the target nucleus is concerned, we will use it as the system
of truly neutral types of neutrons and protons \cite{8}. Our reasoning
refer to those nuclei and leptons, among which no elementary objects
with vector currents.

In the framework of such a presentation, we establish here an explicit mass
structure dependence of leptonic currents of true neutrality. Its analysis
together with earlier measurements \cite{14} of an electric dipole moment
of lepton shows clearly that the existence of axial-vector types of leptons
and fields as well as the availability of mass structure in them are by
no means excluded experimentally.

\begin{center}
{\bf 2. Unity of lepton axial-vector interaction
structural components}
\end{center}

An electroweak scattering of leptons by nuclei in the limit of one-boson
exchange can be described by the two amplitudes \cite{10}:
$$M^{E}_{fi}=\frac{4\pi\alpha}{q_{E}^{2}}\overline{u}(p_{E}',s')\gamma_{5}
\{\gamma_{\mu}[g_{1l^{A}}^{E}(0)+\frac{1}{6}q_{E}^{2}<r^{2}_{l^{A}}>_{E}]-$$
\begin{equation}
-i\sigma_{\mu\lambda}q_{\lambda E}g_{2l^{A}}^{E}(0)\}
u(p_{E},s)<f|J_{\mu}^{\gamma}(q_{E})|i>,
\label{5}
\end{equation}
\begin{equation}
M^{W}_{fi}=
-\frac{G_{F}}{\sqrt{2}}\overline{u}(p_{W}',s')\gamma_{\mu}\gamma_{5}
g_{A_{l}}^{*}u(p_{W},s)<f|J_{\mu}^{Z}(q_{W})|i>.
\label{6}
\end{equation}

Among them $l^{A}=l_{L,R}^{A}(\overline{l}_{R,L}^{A}),$ $q_{E}=p_{E}-p_{E}',$
$q_{W}=p_{W}-p_{W}',$ $p_{E}(p_{W})$ and $p_{E}'(p_{W}')$ denote the
four-momentum of lepton before and after the electric (weak) emission,
$g_{il^{A}}^{E}$ and $<r^{2}_{l^{A}}>_{E}$ characterize a latent Coulomb
$m_{l^{A}}^{E}$ mass dependence of a particle anapole charge, charge radius
and electric dipole moment, $J_{\mu}^{\gamma}$ and $J_{\mu}^{Z}$ describe
the nuclear currents in the interactions with photon and $Z$-boson,
$g_{A_{l}}^{*}$ distinguishes from $g_{A_{l}},$ namely, from the
constant of leptonic axial-vector weak neutral current by a multiplier
$(1/\sin\theta_{W})$ arising at $e_{l^{A}}^{E}=1$ and
\begin{equation}
e_{l^{A}}^{E}=e_{l^{A}}^{W}\sin\theta_{W}.
\label{7}
\end{equation}

In the case of exchange by the two bosons of the Coulomb and weak nature,
only the mixedly interference $(I)$ interaction
$$ReM^{E}_{fi}M^{*W}_{fi}=
-\frac{4\pi\alpha G_{F}}{\sqrt{2}q_{I}^{2}}
Re\Lambda_{I}\Lambda_{I}'\gamma_{5}\{\gamma_{\mu}
[g_{1l^{A}}^{I}(0)+$$
$$+\frac{1}{6}q_{I}^{2}<r^{2}_{l^{A}}>_{I}]-$$
\begin{equation}
-i\sigma_{\mu\lambda}q_{\lambda I}g_{2l^{A}}^{I}(0)\}
\gamma_{\mu}\gamma_{5}g_{A_{l}}^{*}
J_{\mu}^{\gamma}(q_{I})J_{\mu}^{Z}(q_{I})
\label{8}
\end{equation}
is responsible for the investigated process, in which $g_{il^{A}}^{I}$ and
$<r^{2}_{l^{A}}>_{I}$ are present as a consequence of compound structure of
mass and charge. We have also used the notations
$$q_{I}=p_{I}-p_{I}',$$
$$\Lambda_{I}=u(p_{I},s)\overline{u}(p_{I},s),$$
$$\Lambda_{I}'=u(p_{I}',s')\overline{u}(p_{I}',s').$$
Here $p_{I}$ and $p_{I}'$ correspond to the four-momentum of a particle
before and after the electroweakly interference scattering.

This presentation can be explained by the fact that a fermion interference
$(I)$ mass is not equal to its all $(EW)$ rest mass. A distinction in sizes of
both types of masses expresses the unidenticality of a particle mass electric
$(E)$ and weak $(W)$ parts. Such a connection, as well as (\ref{7}), there
arises at the unification of electroweak forces at the more fundamental
dynamical level.

If choose spinless nuclei of the C-noninvariant nature and longitudinal
polarized leptons of true neutrality, the differential cross section of
the considered process at the account of (\ref{5})-(\ref{8}) and
of the standard definition
\begin{equation}
\frac{d\sigma_{EW}(s,s')}{d\Omega}=
\frac{1}{16\pi^{2}}|M^{E}_{fi}+M^{W}_{fi}|^{2}
\label{9}
\end{equation}
can be written in the form
$$d\sigma_{EW}^{A_{l}}(\theta_{EW},s,s')=
d\sigma_{E}^{A_{l}}(\theta_{E},s,s')+$$
\begin{equation}
+d\sigma_{I}^{A_{l}}(\theta_{I},s,s')+
d\sigma_{W}^{A_{l}}(\theta_{W},s,s')
\label{10}
\end{equation}
where $\theta_{EW}$ implies the scattering angle of a studied particle
after an electroweakly united $(EW)$ axial-vector emission.

The first cross section here corresponds to the purely Coulomb
interaction and is equal to
$$\frac{d\sigma_{E}^{A_{l}}(\theta_{E},s,s')}{d\Omega}=
\frac{1}{2}\sigma^{E}_{o}\{(1+ss')[g_{1l^{A}}^{E}-$$
$$-\frac{2}{3}<r^{2}_{l^{A}}>_{E}(m_{l^{A}}^{E})^{2}\gamma_{E}^{-1}]^{2}+$$
\begin{equation}
+4(m_{l^{A}}^{E})^{2}\eta_{E}^{-2}(1-ss')g_{2l^{A}}^{2}
tg^{2}\frac{\theta_{E}}{2}\}F_{E}^{2}(q_{E}^{2}).
\label{11}
\end{equation}

The second term is explained by the electroweakly mixed
interaction (\ref{8}) and has the form
$$\frac{d\sigma_{I}^{A_{l}}(\theta_{I},s,s')}{d\Omega}=
\frac{1}{2}\rho_{I}\sigma^{I}_{o}
g_{A_{l}}(1+ss')\{g_{1l^{A}}^{I}-$$
\begin{equation}
-\frac{2}{3}<r^{2}_{l^{A}}>_{I}(m_{l^{A}}^{I})^{2}
\gamma_{I}^{-1}\}F_{I}(q_{I}^{2}).
\label{12}
\end{equation}

The weak scattering can be described by the cross section
$$\frac{d\sigma_{W}^{A_{l}}(\theta_{W},s,s')}{d\Omega}=
\frac{G_{F}^{2}(m_{l^{A}}^{W})^{2}}{16\pi^{2}\sin^{2}\theta_{W}}
g_{A_{l}}^{2}\eta_{W}^{-2}(1+ss')[1-$$
\begin{equation}
-\eta_{W}^{2}]F_{W}^{2}(q_{W}^{2})
\cos^{2}\frac{\theta_{W}}{2}.
\label{13}
\end{equation}
Here it has been accepted that
$$\sigma_{o}^{E}=\frac{\alpha^{2}}{4(m_{l^{A}}^{E})^{2}}
\frac{\gamma_{E}^{2}}{\alpha_{E}}, \, \, \, \,
\rho_{I}=-\frac{2G_{F}(m_{l^{A}}^{I})^{2}}
{\pi\sqrt{2}\alpha\sin\theta_{W}}\gamma_{I}^{-1},$$
$$\sigma_{o}^{I}=\frac{\alpha^{2}}{4(m_{l^{A}}^{I})^{2}}
\frac{\gamma_{I}^{2}}{\alpha_{I}}, \, \, \, \,
\alpha_{K}=\frac{\eta^{2}_{K}}{(1-\eta^{2}_{K})
\cos^{2}(\theta_{K}/2)},$$
$$\gamma_{K}=\frac{\eta^{2}_{K}}{(1-\eta^{2}_{K})
\sin^{2}(\theta_{K}/2)}, \, \, \, \,
\eta_{K}=\frac{m_{l^{A}}^{K}}{E_{l^{A}}^{K}},$$
$$F_{E}(q_{E}^{2})=ZF_{c}(q_{E}^{2}), \, \, \, \,
F_{I}(q_{I}^{2})=ZZ_{W}F_{c}^{2}(q_{I}^{2}),$$
$$F_{W}(q_{W}^{2})=Z_{W}F_{c}(q_{W}^{2}), \, \, \, \,
q_{K}^{2}=-4(m_{l^{A}}^{K})^{2}\gamma_{K}^{-1},$$
$$Z_{W}=\frac{1}{2}\{\beta_{A}^{(0)}(Z+N)+\beta_{A}^{(1)}(Z-N)\},$$
$$A=Z+N, \, \, \, \, M_{T}=\frac{1}{2}(Z-N),$$
$$\beta_{A}^{(0)}=-2\sin^{2}\theta_{W},
\, \, \, \, \beta_{A}^{(1)}=\frac{1}{2}-2\sin^{2}\theta_{W},$$
$$g_{A_{l}}=-\frac{1}{2}, \, \, \, \, K=E,I,W.$$

Of them $\theta_{K}$ characterize the lepton Coulomb, weak and interference
scattering angles at the energies
$$E_{l^{A}}^{K}=\sqrt{p_{K}^{2}+(m_{l^{A}}^{K})^{2}},$$
the functions $F_{c}(q_{K}^{2})$ describe in these three types of processes
the charge $(F_{c}(0)=1)$ form factors of an electroweakly charged C-odd
nucleus with the projection $M_{T}$ of its isospin $T,$ $\beta_{A}^{(0)}$
and $\beta_{A}^{(1)}$ denote herewith the isoscalar and isovector component
constants of the used nucleus axial-vector weak neutral current.

The cross section of Coulomb scattering consists of the purely self
interference parts $(g_{il^{A}}^{E})^{2}$ and $<r_{l^{A}}^{4}>_{E}$ including
as well as the contribution $g_{1l^{A}}^{E}<r_{l^{A}}^{2}>_{E}$ of the mixed
interference between the interactions of the anapole charge and its radius
with a nucleus. Such components of the formula (\ref{11}) can explain the
formation of the left - or right-handed \cite{5} dileptons of
axial-vector currents:
\begin{equation}
(l^{A}_{L}, \overline{l}_{R}^{A}), \, \, \, \,
(l^{A}_{R}, \overline{l}_{L}^{A}).
\label{14}
\end{equation}

In the presence of the interaction (\ref{8}), each of these types of
paraparticles requires the appearance in (\ref{12}) of one of its structural
parts $g_{A_{l}}g_{1l^{A}}^{I}$ or $g_{A_{l}}<r_{l^{A}}^{2}>_{I}$ and thereby
confirms the fact that the structure of difermions is defined by the behavior
of their interaction. Therefore, the self interference contributions
$g_{A_{l}}^{2}$ in (\ref{13}) can be considered as an indication to
the existence of paraleptons of the weak neutral $A_{l}$ currents.

We have already mentioned that (\ref{9}) redoubles the mixedly interference
cross sections. However, the number of difermions does not distinguish from
the number of those processes, in which they can appear. Such a conformity
is responsible for separation of any type of the mixedly interference
component of the interaction cross section into the two.

By following the structure of formulas (\ref{11})-(\ref{13}), it is easy to
observe the terms $(1+ss')$ and $(1-ss')$ which say in favor of the scattering
with $(s'=s)$ or without $(s'=-s)$ change of helicities of the left
$(s=-1)$ - and right $(s=+1)$-handed leptons of true neutrality. Therefore,
if we sum each of them over $s',$ we can present (\ref{10}) in the form
$$d\sigma_{EW}^{A_{l}}(\theta_{EW},s)=d\sigma_{E}^{A_{l}}(\theta_{E},s)+
\frac{1}{2}d\sigma_{I}^{A_{l}}(\theta_{I},s)+$$
\begin{equation}
+\frac{1}{2}d\sigma_{I}^{A_{l}}(\theta_{I},s)+
d\sigma_{W}^{A_{l}}(\theta_{W},s)
\label{15}
\end{equation}
where the purely Coulomb process cross section has the size
$$d\sigma_{E}^{A_{l}}(\theta_{E},s)=
d\sigma_{E}^{A_{l}}(\theta_{E},g_{1l^{A}}^{E},s)+\frac{1}{2}
d\sigma_{E}^{A_{l}}(\theta_{E},g_{1l^{A}}^{E},<r^{2}_{l^{A}}>_{E},s)+$$
$$+\frac{1}{2}
d\sigma_{E}^{A_{l}}(\theta_{E},g_{1l^{A}}^{E},<r^{2}_{l^{A}}>_{E},s)+$$
\begin{equation}
+d\sigma_{E}^{A_{l}}(\theta_{E},<r^{2}_{l^{A}}>_{E},s)+
d\sigma_{E}^{A_{l}}(\theta_{E},g_{2l^{A}}^{E},s),
\label{16}
\end{equation}
\begin{equation}
\frac{d\sigma_{E}^{A_{l}}(\theta_{E},g_{1l^{A}}^{E},s)}{d\Omega}=
\frac{d\sigma_{E}^{A_{l}}(\theta_{E},g_{1l^{A}}^{E},s'=s)}{d\Omega}=
\sigma^{E}_{o}(g_{1l^{A}}^{E})^{2}F_{E}^{2}(q^{2}_{E}),
\label{17}
\end{equation}

$$\frac{d\sigma_{E}^{A_{l}}(\theta_{E},g_{1l^{A}}^{E},<r^{2}_{l^{A}}>_{E},s)}
{d\Omega}=
\frac{d\sigma_{E}^{A_{l}}(\theta_{E},g_{1l^{A}}^{E},<r^{2}_{l^{A}}>_{E},s'=s)}
{d\Omega}=$$
\begin{equation}
=-\frac{2}{3}(m_{l^{A}}^{E})^{2}\gamma_{E}^{-1}\sigma^{E}_{o}
g_{1l^{A}}^{E}<r^{2}_{l^{A}}>_{E}F_{E}^{2}(q^{2}_{E}),
\label{18}
\end{equation}

$$\frac{d\sigma_{E}^{A_{l}}(\theta_{E},<r^{2}_{l^{A}}>_{E},s)}{d\Omega}=
\frac{d\sigma_{E}^{A_{l}}(\theta_{E},<r^{2}_{l^{A}}>_{E},s'=s)}{d\Omega}=$$
\begin{equation}
=\frac{4}{9}(m_{l^{A}}^{E})^{4}\gamma_{E}^{-2}\sigma^{E}_{o}
<r^{4}_{l^{A}}>_{E}F_{E}^{2}(q^{2}_{E}),
\label{19}
\end{equation}

$$\frac{d\sigma_{E}^{A_{l}}(\theta_{E},g_{2l^{A}}^{E},s)}{d\Omega}=
\frac{d\sigma_{E}^{A_{l}}(\theta_{E},g_{2l^{A}}^{E},s'=-s)}{d\Omega}=$$
\begin{equation}
=4(m_{l^{A}}^{E})^{2}\eta^{-2}_{E}\sigma^{E}_{o}
(g_{2l^{A}}^{E})^{2}F_{E}^{2}(q^{2}_{E})tg^{2}\frac{\theta_{E}}{2}.
\label{20}
\end{equation}

The second cross section is responsible for the electroweakly interference
scattering and behaves as
$$d\sigma_{I}^{A_{l}}(\theta_{I},s)=
d\sigma_{I}^{A_{l}}(\theta_{I},g_{A_{l}},g_{1l^{A}}^{I},s)+$$
\begin{equation}
+d\sigma_{I}^{A_{l}}(\theta_{I},g_{A_{l}},<r^{2}_{l^{A}}>_{I},s),
\label{21}
\end{equation}

$$\frac{d\sigma_{I}^{A_{l}}(\theta_{I},g_{A_{l}},g_{1l^{A}}^{I},s)}{d\Omega}=
\frac{d\sigma_{I}^{A_{l}}(\theta_{I},g_{A_{l}},g_{1l^{A}}^{I},s'=s)}
{d\Omega}=$$
\begin{equation}
=\rho_{I}\sigma^{I}_{o}g_{A_{l}}g_{1l^{A}}^{I}F_{I}(q_{I}^{2}),
\label{22}
\end{equation}

$$\frac{d\sigma_{I}^{A_{l}}(\theta_{I},g_{A_{l}},<r^{2}_{l^{A}}>_{I},s)}
{d\Omega}=
\frac{d\sigma_{I}^{A_{l}}(\theta_{I},g_{A_{l}},<r^{2}_{l^{A}}>_{I},s'=s)}
{d\Omega}+$$
\begin{equation}
=-\frac{2}{3}(m_{l^{A}}^{I})^{2}\gamma_{I}^{-1}
\rho_{I}\sigma^{I}_{o}g_{A_{l}}<r^{2}_{l^{A}}>_{I}F_{I}(q_{I}^{2}).
\label{23}
\end{equation}

Third term corresponds to purely weak process and is equal to
$$\frac{d\sigma_{W}^{A_{l}}(\theta_{W},g_{A_{l}},s)}{d\Omega}=
\frac{d\sigma_{W}^{A_{l}}(\theta_{W},g_{A_{l}},s'=s)}{d\Omega}=$$
\begin{equation}
=\frac{G_{F}^{2}(m_{l^{A}}^{W})^{2}}{8\pi^{2}\sin^{2}\theta_{W}}
g_{A_{l}}^{2}\eta_{W}^{-2}(1-\eta_{W}^{2})
F_{W}^{2}(q_{W}^{2})\cos^{2}\frac{\theta_{W}}{2}.
\label{24}
\end{equation}

If compare (\ref{16})-(\ref{24}), one can see that (\ref{18}) and (\ref{23})
have the negative signs. This testifies in favor of coexistence of the anapole
$g_{1l^{A}}$ charge of lepton and its charge $r_{l^{A}}$ radius. On the other
hand, as was noted in the work \cite{3}, between $g_{1l^{A}}$ and $g_{2l^{A}}$
there exists a hard connection. In such a situation, appears the regularity
that at the availability of an axial-vector mass, any of truly neutral leptons
can possess simultaneously each of these three types of $A_{l}$ currents.

It is desirable to also replace (\ref{10}) after averaging
(\ref{11})-(\ref{13}) over $s$ and summing over $s'$ for
$$d\sigma_{EW}^{A_{l}}(\theta_{EW})=d\sigma_{E}^{A_{l}}(\theta_{E})+
\frac{1}{2}d\sigma_{I}^{A_{l}}(\theta_{I})+$$
\begin{equation}
+\frac{1}{2}d\sigma_{I}^{A_{l}}(\theta_{I})+
d\sigma_{W}^{A_{l}}(\theta_{W}).
\label{25}
\end{equation}

Similarly to the structural parts of (\ref{15}), any cross section
here has the self value:
$$d\sigma_{E}^{A_{l}}(\theta_{E})=
d\sigma_{E}^{A_{l}}(\theta_{E},g_{1l^{A}}^{E})+
\frac{1}{2}
d\sigma_{E}^{A_{l}}(\theta_{E},g_{1l^{A}}^{E},<r^{2}_{l^{A}}>_{E})+$$
$$+\frac{1}{2}
d\sigma_{E}^{A_{l}}(\theta_{E},g_{1l^{A}}^{E},<r^{2}_{l^{A}}>_{E})+$$
\begin{equation}
+d\sigma_{E}^{A_{l}}(\theta_{E},<r^{2}_{l^{A}}>_{E})+
d\sigma_{E}^{A_{l}}(\theta_{E},g_{2l^{A}}^{E}),
\label{26}
\end{equation}
\begin{equation}
d\sigma_{I}^{A_{l}}(\theta_{I})=
d\sigma_{I}^{A_{l}}(\theta_{I},g_{A_{l}},g_{1l^{A}}^{I})+
d\sigma_{I}^{A_{l}}(\theta_{I},g_{A_{l}},<r^{2}_{l^{A}}>_{I}),
\label{27}
\end{equation}
\begin{equation}
d\sigma_{W}^{A_{l}}(\theta_{W})=
d\sigma_{W}^{A_{l}}(\theta_{W},g_{A_{l}}).
\label{28}
\end{equation}

Each component of any of them is equal to the corresponding contribution
from (\ref{16}), (\ref{21}), (\ref{24}) and that, consequently, there
are the ratios
\begin{equation}
d\sigma_{E}^{A_{l}}(\theta_{E},g_{1l^{A}}^{E})=
d\sigma_{E}^{A_{l}}(\theta_{E},g_{1l^{A}}^{E},s),
\label{29}
\end{equation}
\begin{equation}
d\sigma_{E}^{A_{l}}(\theta_{E},g_{1l^{A}}^{E},<r^{2}_{l^{A}}>_{E})=
d\sigma_{E}^{A_{l}}(\theta_{E},g_{1l^{A}}^{E},<r^{2}_{l^{A}}>_{E},s),
\label{30}
\end{equation}
\begin{equation}
d\sigma_{E}^{A_{l}}(\theta_{E},<r^{2}_{l^{A}}>_{E})=
d\sigma_{E}^{A_{l}}(\theta_{E},<r^{2}_{l^{A}}>_{E},s),
\label{31}
\end{equation}
\begin{equation}
d\sigma_{E}^{A_{l}}(\theta_{E},g_{2l^{A}}^{E})=
d\sigma_{E}^{A_{l}}(\theta_{E},g_{2l^{A}}^{E},s),
\label{32}
\end{equation}
\begin{equation}
d\sigma_{I}^{A_{l}}(\theta_{I},g_{A_{l}},g_{1l^{A}}^{I})=
d\sigma_{I}^{A_{l}}(\theta_{I},g_{A_{l}},g_{1l^{A}}^{I},s),
\label{33}
\end{equation}
\begin{equation}
d\sigma_{I}^{A_{l}}(\theta_{I},g_{A_{l}},<r^{2}_{l^{A}}>_{I})=
d\sigma_{I}^{A_{l}}(\theta_{I},g_{A_{l}},<r^{2}_{l^{A}}>_{I},s),
\label{34}
\end{equation}
\begin{equation}
d\sigma_{W}^{A_{l}}(\theta_{W},g_{A_{l}})=
d\sigma_{W}^{A_{l}}(\theta_{W},g_{A_{l}},s).
\label{35}
\end{equation}

The above expansions (\ref{15}) and (\ref{25}) would seem to say about
that either incoming leptons are strictly unpolarized or they have the left
(right)-handed longitudinal polarization. One can convince, however, that this
is not quite so. The point is that the united structure of medium \cite{15}
where a particle interacts with a nucleus can essentially influence on its
spin properties. At the same time, the scattering itself becomes possible
owing to the unified nature of the anapole and electric dipole. The scattered
leptons can therefore be partially ordered class of outgoing fermions.

Thus, (\ref{10}) constitutes a set of the scattering cross section
of partially ordered flux of axial-vector particles
\begin{equation}
d\sigma_{EW}^{A_{l}}=\{d\sigma_{EW}^{A_{l}}(\theta_{EW},s), \, \, \, \,
d\sigma_{EW}^{A_{l}}(\theta_{EW})\}.
\label{36}
\end{equation}

It shows that each of (\ref{15}) and (\ref{25}) behaves as a subclass:
$$d\sigma_{EW}^{A_{l}}(\theta_{EW},s)=
\{d\sigma_{E}^{A_{l}}(\theta_{E},g_{1l^{A}}^{E},s),  \, \, \, \,
\frac{1}{2}
d\sigma_{E}^{A_{l}}(\theta_{E},g_{1l^{A}}^{E},<r^{2}_{l^{A}}>_{E},s),$$
$$\frac{1}{2}
d\sigma_{E}^{A_{l}}(\theta_{E},g_{1l^{A}}^{E},<r^{2}_{l^{A}}>_{E},s),
\, \, \, \,
d\sigma_{E}^{A_{l}}(\theta_{E},<r^{2}_{l^{A}}>_{E},s),$$
$$d\sigma_{E}^{A_{l}}(\theta_{E},g_{2l^{A}}^{E},s),
\, \, \, \,
\frac{1}{2}
d\sigma_{I}^{A_{l}}(\theta_{I},g_{A_{l}},g_{1l^{A}}^{I},s),$$
$$\frac{1}{2}
d\sigma_{I}^{A_{l}}(\theta_{I},g_{A_{l}},g_{1l^{A}}^{I},s),
\, \, \, \,
\frac{1}{2}
d\sigma_{I}^{A_{l}}(\theta_{I},g_{A_{l}},<r^{2}_{l^{A}}>_{I},s),$$
\begin{equation}
\frac{1}{2}
d\sigma_{I}^{A_{l}}(\theta_{I},g_{A_{l}},<r^{2}_{l^{A}}>_{I},s),
\, \, \, \,
d\sigma_{W}^{A_{l}}(\theta_{W},g_{A_{l}},s)\},
\label{37}
\end{equation}
$$d\sigma_{EW}^{A_{l}}(\theta_{EW})=
\{d\sigma_{E}^{A_{l}}(\theta_{E},g_{1l^{A}}^{E}),
\, \, \, \,
\frac{1}{2}
d\sigma_{E}^{A_{l}}(\theta_{E},g_{1l^{A}}^{E},<r^{2}_{l^{A}}>_{E}),$$
$$\frac{1}{2}
d\sigma_{E}^{A_{l}}(\theta_{E},g_{1l^{A}}^{E},<r^{2}_{l^{A}}>_{E}),
\, \, \, \,
d\sigma_{E}^{A_{l}}(\theta_{E},<r^{2}_{l^{A}}>_{E}),$$
$$d\sigma_{E}^{A_{l}}(\theta_{E},g_{2l^{A}}^{E}), \, \, \, \,
\frac{1}{2}
d\sigma_{I}^{A_{l}}(\theta_{I},g_{A_{l}},g_{1l^{A}}^{I}),$$
$$\frac{1}{2}
d\sigma_{I}^{A_{l}}(\theta_{I},g_{A_{l}},g_{1l^{A}}^{I}),
\, \, \, \,
\frac{1}{2}
d\sigma_{I}^{A_{l}}(\theta_{I},g_{A_{l}},<r^{2}_{l^{A}}>_{I}),$$
\begin{equation}
\frac{1}{2}
d\sigma_{I}^{A_{l}}(\theta_{I},g_{A_{l}},<r^{2}_{l^{A}}>_{I}),
\, \, \, \,
d\sigma_{W}^{A_{l}}(\theta_{W},g_{A_{l}})\}.
\label{38}
\end{equation}

It is seen from (\ref{29})-(\ref{35}) that this subsets have the same size.
Such a possibility is realized only if cross sections (\ref{15}) and (\ref{25})
are responsible for those processes which can originate as a consequence of
formation of the same difermions of an axial-vector nature. In other words,
all elements of sets (\ref{37}) and (\ref{38}) correspond to one of the
above-mentioned types of spin states of paraleptons. In this appears an
equality of the investigated process cross section structural parts.

At the same time, it is clear that between the leptons of any of parafermions
(\ref{14}) there exists a flavour symmetrical connection \cite{3}. Such
a unification gives the right to interpret the flavour symmetry as a
theorem \cite{5} about an equality of the structural components of lepton
interaction cross sections with an electroweak field of true neutrality.

\begin{center}
{\bf 3. Connection between leptonic currents of an axial-vector nature}
\end{center}

We have already seen that the ratio of each pair of elements in (\ref{36})
is equal to unity. Such an equality can constitute a set of the forty two
relations. They together with sizes of cross sections (\ref{29})-(\ref{35}),
establish the system of the twenty one explicit equations.

For elucidation of their ideas, it is desirable to choose the five of them:
\begin{equation}
\frac{d\sigma_{E}^{A_{l}}(\theta_{E},<r^{2}_{l^{A}}>_{E})}
{d\sigma_{E}^{A_{l}}(\theta_{E},g_{il^{A}}^{E})}=1,
\, \, \, \,
\frac{d\sigma_{I}^{A_{l}}(\theta_{I},g_{A_{l}},g_{1l^{A}}^{I})}
{2d\sigma_{E}^{A_{l}}(\theta_{E},g_{1l^{A}}^{E})}=1,
\label{39}
\end{equation}
\begin{equation}
\frac{2d\sigma_{W}^{A_{l}}(\theta_{W},g_{A_{l}})}
{d\sigma_{E}^{A_{l}}(\theta_{E},g_{1l^{A}}^{E},<r^{2}_{l^{A}}>_{E})}=1,
\, \, \, \,
\frac{d\sigma_{I}^{A_{l}}(\theta_{I},g_{A_{l}},<r^{2}_{l^{A}}>_{I})}
{d\sigma_{I}^{A_{l}}(\theta_{I},g_{A_{l}},g_{1l^{A}}^{I})}=1.
\label{40}
\end{equation}

Insertion of the corresponding values of cross sections from
(\ref{29})-(\ref{35}) in (\ref{39}) and (\ref{40}) jointly with the limits
$$lim_{\eta_{E}\rightarrow 0,\theta_{E}\rightarrow 0}
\frac{tg^{2}(\theta_{E}/2)}{\eta_{E}^{2}}=\frac{1}{4},$$
$$lim_{\eta_{K}\rightarrow 0,\theta_{K}\rightarrow 0}
\frac{\eta^{2}_{K}}{(1-\eta^{2}_{K})
\sin^{2}(\theta_{K}/2)}=-2$$
leads us once again to the system
\begin{equation}
\frac{1}{3}<r^{2}_{l^{A}}>_{E}(m_{l^{A}}^{E})^{2}=\pm g_{1l^{A}}^{E}(0),
\label{41}
\end{equation}
\begin{equation}
\frac{1}{3}<r^{2}_{l^{A}}>_{E}m_{l^{A}}^{E}=\pm g_{2l^{A}}^{E}(0),
\label{42}
\end{equation}
\begin{equation}
(g_{1l^{A}}^{E}(0))^{2}=g_{A_{l}}\frac{G_{F}(m_{l^{A}}^{E})^{2}}
{2\pi\sqrt{2}\alpha}\frac{Z_{W}^{*}}{\sin\theta_{W}}g_{1l^{A}}^{I}(0),
\label{43}
\end{equation}
\begin{equation}
\frac{1}{3}<r^{2}_{l^{A}}>_{E}g_{1l^{A}}^{E}(0)=g_{A_{l}}^{2}
\frac{G_{F}^{2}(m_{l^{A}}^{W})^{2}}{8\pi^{2}\alpha^{2}}
\left(\frac{Z_{W}^{*}}{\sin\theta_{W}}\right)^{2},
\label{44}
\end{equation}
\begin{equation}
\frac{1}{3}<r^{2}_{l^{A}}>_{I}(m_{l^{A}}^{I})^{2}=g_{1l^{A}}^{I}(0).
\label{45}
\end{equation}

The availability of a multiplier $Z_{W}^{*}=Z_{W}/Z$ in (\ref{43}) and
(\ref{44}) implies the possible change of the innate sizes of currents
$g_{il^{A}}^{E},$ $<r^{2}_{l^{A}}>_{E},$ $g_{1l^{A}}^{I}$ and
$<r^{2}_{l^{A}}>_{I}$ in the result of influence of the target nucleus
isotopic structure on these properties of particles \cite{16}.

\begin{center}
{\bf 4. Latent structure of axial-vector leptons}
\end{center}

At the choice of a nucleus with the same number $(Z=N)$ of protons and
neutrons, the solution of equations (\ref{41})-(\ref{45}) defines an explicit
mass structure dependence of leptonic currents of an axial-vector nature
\begin{equation}
g_{1l^{A}}^{E}(0)=-g_{A_{l}}
\frac{G_{F}m_{l^{A}}^{E}m_{l^{A}}^{W}}{\pi\sqrt{2}\alpha}\sin\theta_{W},
\label{46}
\end{equation}
\begin{equation}
g_{2l^{A}}^{E}(0)=-g_{A_{l}}
\frac{G_{F}m_{l^{A}}^{W}}{\pi\sqrt{2}\alpha}\sin\theta_{W},
\label{47}
\end{equation}
\begin{equation}
<r^{2}_{l^{A}}>_{E}=-g_{A_{l}}
\frac{3G_{F}}{\pi\sqrt{2}\alpha}
\frac{m_{l^{A}}^{W}}{m_{l^{A}}^{E}}\sin\theta_{W},
\label{48}
\end{equation}
\begin{equation}
g_{1l^{A}}^{I}(0)=-g_{A_{l}}
\frac{G_{F}(m_{l^{A}}^{W})^{2}}{\pi\sqrt{2}\alpha}\sin\theta_{W},
\label{49}
\end{equation}
\begin{equation}
<r^{2}_{l^{A}}>_{I}=-g_{A_{l}}\frac{3G_{F}}{\pi\sqrt{2}\alpha}
\left(\frac{m_{l^{A}}^{W}}{m_{l^{A}}^{I}}\right)^{2}\sin\theta_{W}.
\label{50}
\end{equation}

It is clear, however, that
\begin{equation}
e_{l^{A}}^{E}=-g_{A_{l}}
\frac{G_{F}m_{l^{A}}^{E}m_{l^{A}}^{W}}{\pi\sqrt{2}\alpha}\sin\theta_{W}
\label{51}
\end{equation}
coincides with the renormalized size of C-noninvariant electric charge. In
this situation, (\ref{46})-(\ref{50}) can be expressed in a latent united form
\begin{equation}
g_{1l^{A}}^{E}(0)=e_{l^{A}}^{E},
\label{52}
\end{equation}
\begin{equation}
g_{2l^{A}}^{E}(0)=\frac{e_{l^{A}}^{E}}{m_{l^{A}}^{E}},
\label{53}
\end{equation}
\begin{equation}
<r^{2}_{l^{A}}>_{E}=\frac{3e_{l^{A}}^{E}}{(m_{l^{A}}^{E})^{2}},
\label{54}
\end{equation}
\begin{equation}
g_{1l^{A}}^{I}(0)=\frac{m_{l^{A}}^{W}}{m_{l^{A}}^{E}}e_{l^{A}}^{E},
\label{55}
\end{equation}
\begin{equation}
<r^{2}_{l^{A}}>_{I}=
\frac{m_{l^{A}}^{E}m_{l^{A}}^{W}}{(m_{l^{A}}^{I})^{2}}<r^{2}_{l^{A}}>_{E}.
\label{56}
\end{equation}

They show that $g_{2l^{A}}^{E}$ corresponds to a Dirac value of the electric
dipole. Insofar as its anomalous component is concerned, it can essentially
appear at the exchange by the two bosons.

\begin{center}
{\bf 5. Conclusion}
\end{center}

Already known laboratory facts \cite{14} for the lepton electric dipole
moment give the following estimates:
\begin{equation}
d_{e^{A}}^{E}=0.07\cdot 10^{-26}\ {\rm e\cdot cm},
\label{57}
\end{equation}
\begin{equation}
d_{\mu^{A}}^{E}=3.7\cdot 10^{-19}\ {\rm e\cdot cm},
\label{58}
\end{equation}
\begin{equation}
d_{\tau^{A}}^{E}=0.45\cdot 10^{-16}\ {\rm e\cdot cm}.
\label{59}
\end{equation}

In establishing these sizes, it has been usually assumed that $e$ is equal
to an electric $e_{l^{V}}^{E}$ charge of lepton of a vector nature. If we
accept this idea, comparing (\ref{53}) with (\ref{57})-(\ref{59}), we would
define the other values of C-invariant lepton masses instead of the available
data \cite{14} in the literature. They satisfy the inequalities
\begin{equation}
e_{l^{A}}^{K}\neq e_{l^{V}}^{K}, \, \, \, \,
m_{l^{A}}^{K}\neq m_{l^{V}}^{K}
\label{60}
\end{equation}
which follow from the fact that usual Dirac fermions having the mass of a
vector character \cite{3} possess no an axial-vector interaction \cite{5}.

Turning to (\ref{52}) and (\ref{53}), we remark that
\begin{equation}
\frac{g_{1l^{A}}^{E}(0)}{g_{2l^{A}}^{E}(0)}=m_{l^{A}}^{E}.
\label{61}
\end{equation}

It is compatible with that take places at the availability of the same
electric charge of an axial-vector character in all types
of C-noninvariant leptons.

Another serious basis for such a statement is that flavour symmetry expresses
the lepton universality \cite{8}. This principle is a general and does not
depend of whether the fermions have a vector or an axial-vector interaction.
Thereby, it requires the establishment of nature of truly neutral types of
leptons and fields from the point of view of constancy law of the size
\begin{equation}
m_{l^{A}}^{E}m_{l^{A}}^{W}=const.
\label{62}
\end{equation}

According to one of its aspects, all C-noninvariant leptons regardless
of the difference in masses of an axial-vector character, possess an equal
anapole charge with his radius as well as an identical electric dipole moment.

From our earlier developments \cite{17}, we find that the standard formula
for $e_{l^{V}}^{E}$ with the use of a relation
\begin{equation}
\frac{G_{F}}{\pi\sqrt{2}\alpha}=
\frac{1}{2m_{W}^{2}}\frac{1}{\sin^{2}\theta_{W}}
\label{63}
\end{equation}
can be presented as
\begin{equation}
e_{l^{V}}^{E}=-g_{V_{l}}\frac{m_{l^{V}}^{E}m_{l^{V}}^{W}}
{2m_{W^{\pm}}^{2}}\frac{1}{\sin\theta_{W}}.
\label{64}
\end{equation}

The latter would seem to indicate that (\ref{63}) gives the possibility to
express the size of (\ref{46}) through the charge of C-invariant lepton. On
the other hand, such a unification of (\ref{46}), (\ref{47}) and (\ref{64})
together with (\ref{57})-(\ref{59}) allows to establish those values of the
weak axial-vector masses, at which all predictions of the universality
law (\ref{62}) say in favor of that truly neutral leptons due to their
C-noninvariant nature possess no interaction with $W^{\pm}$-bosons. They
have with $W^{0}$-bosons the same interactions as the nucleons of
an axial-vector character.

We can, therefore, reduce (\ref{51}) to the following:
\begin{equation}
e_{l^{A}}^{E}=-g_{A_{l}}\frac{m_{l^{A}}^{E}m_{l^{A}}^{W}}
{2m_{W^{0}}^{2}}\frac{1}{\sin\theta_{W}}.
\label{65}
\end{equation}

This can be based on the fact that C-even or C-odd fermion constitutes,
respectively, a vector or an axial-vector gauge boson.

Thus, we have established a part of nature of elementary particles in which
there exist the massive $Z^{\pm} (W^{0})$-bosons. From their point of view,
the interaction between the C-symmetrical (C-antisymmetrical) lepton and
field of emission there arises at the expense of exchange not only by vector
(axial-vector) photon but also by $Z^{\pm} (Z^{0})$-boson. Herewith a group
is obtained of arguments in favor of the unidenticality of masses
\begin{equation}
m_{Z^{0}}\neq m_{Z^{\pm}},
\label{66}
\end{equation}
\begin{equation}
m_{W^{0}}\neq m_{W^{\pm}}
\label{67}
\end{equation}
which in conformity with the ideas \cite{18} of the standard electroweak
theory state that
\begin{equation}
m_{W^{0}}=m_{Z^{0}}\cos\theta_{W}.
\label{68}
\end{equation}
\begin{equation}
m_{W^{\pm}}=m_{Z^{\pm}}\cos\theta_{W}.
\label{69}
\end{equation}

So, we must recognize that all earlier measured properties of gauge
bosons $Z (W)$ can be interpreted as those features of fields which refer
undoubtedly only to $Z^{\pm} (W^{\pm})$-bosons. Their nature has been created
so that to any type of a vector particle $Z^{\pm} (W^{\pm})$ corresponds a kind
of axial-vector $Z^{0} (W^{0})$-boson. They constitute herewith the triplets
$(Z^{-},$ $Z^{0},$ $Z^{+})$ and $(W^{-},$ $W^{0},$ $W^{+})$ as a consequence
of a connection between the masses of each of them.

However, in spite of that the above-mentioned triboson unification needs in
special verification, a system of inequalities of (\ref{60}), (\ref{66}) and
(\ref{67}) gives the possibility to understand not only the existence of truly
neutral types of Dirac fermions and gauge fields but also the availability of
mass structure in them as the one of experimentally known facts.

\end{document}